\documentclass{aa}
\usepackage{graphicx}

\newcommand{\HI}{{\ion{H}{i}}}

\newcommand{\abHI}{\rm H{\hskip 0.05cm\scriptsize I}}

\newcommand{\kms}{$\,$km$\,$s$^{-1}$}

\newcommand{\mJybeam}{mJy beam$^{-1}$}

\def\emph#1{{\sl #1}}
\newcommand{\ltsima} {$\; \buildrel < \over \sim \;$}
\newcommand{\gtsima} {$\; \buildrel > \over \sim \;$}
\newcommand{\lta} {\lower.5ex\hbox{\ltsima}}
\newcommand{\gta} {\lower.5ex\hbox{\gtsima}}

\input{psfig}

\begin{document}

\title{Fast neutral outflows in powerful radio galaxies: a major source of
feedback in massive galaxies \thanks{Based on observations
with the Westerbork Synthesis Radio Telescope.}}

\titlerunning{Broad \abHI\ absorption}
\authorrunning{Morganti et al. }

\author{R. Morganti\inst{1,2}, C. N. Tadhunter\inst{3}, T. A. Oosterloo\inst{1,2},
}

\offprints{morganti@astron.nl}

\institute{Netherlands Foundation for Research in Astronomy, Postbus 2,
NL-7990 AA, Dwingeloo, The Netherlands
\and
Kapteyn Astronomical Institute, University of Groningen, P.O. Box 800,
9700 AV Groningen, the Netherlands
\and 
Dep. Physics and Astronomy,
University of Sheffield, Sheffield, S7 3RH, UK
}

\date{Received ...; accepted ...}

\abstract{ We report the detection of fast ($\sim 1000$ \kms), massive
outflows of neutral gas observed -- using the WSRT -- as 21-cm \HI\ absorption
against the strong radio continuum of seven radio sources.  The neutral
outflows occur, in at least somes cases, at kpc distance from the nucleus, and
they are most likely driven by the interactions between the expanding radio
jets and the gaseous medium enshrouding the central regions.  We estimate that
the associated mass outflow rates are up to $\sim 50$ M$_\odot$
yr$^{-1}$, comparable (although at the lower end of the distribution) to the
outflow rates found for starburst-driven superwinds in Ultra Luminous IR
Galaxies (ULIRG).  This suggests that massive, jet-driven outflows of
neutral gas in radio-loud AGN can have as large an impact on the evolution of
the host galaxies as the outflows associated with starbursts.  A radio-loud
phase of the AGN is likely a relatively common, albeit short, phase in the
life of many (or even all) massive ellipticals.  Jet-driven neutral outflows
may represent one of the main feedback mechanisms in these galaxies. 

}
\maketitle

\section{Introduction: gaseous outflows in AGN}

Galactic mass outflows are spectacular events widely recognised to
dramatically affect the evolution of galaxies because of the large amounts of
energy they feed back into the interstellar medium.  They also appear to be a
widespread phenomenon in galaxies, both in the local and in the distant
Universe (Heckman 2002, Rupke et al. 2002, 2005a,b, Veilleux et al. 2005, Frey et
al. 2002). They can be driven by super-winds associated with large starbursts,
or by the energy released by active galactic nuclei (AGN). They have a wide
range of effects, from clearing the circum-nuclear regions and halting the
growth of the supermassive black-holes (Silk \& Rees 1998, Di Matteo et
al. 2005) to injecting energy and metals into the interstellar and
intergalactic medium. Indeed, it has been suggested that AGN-driven outflows
are a major source of feedback in the overall galaxy formation process,
regulating the correlations found between the masses of the central
super-massive black-hole and the properties of the host galaxies, and
preventing the formation of too many massive galaxies in the early universe.
Although numerical simulations demonstrate that AGN-induced outflows
are potentially a key component in the evolution of massive galaxies (Di
Matteo et al. 2005), there remains a dearth of hard observational evidence for
such outflows and the impact they have on the interstellar medium on the
scales of galaxy bulges.  Most of the existing evidence for outflows
associated with AGN has been obtained using X-ray and UV absorption line
observations of the highly ionized gas close to the nucleus, which reveal the
presence of broad and narrow absorption line systems that are significantly
blueshifted (Crenshaw et al. 2003). However, estimates of scales and physical
conditions of such ionised gas outflows are necessarily indirect, their mass
outflow rates are difficult to determine, and it is not yet clear whether they
are a significant source of feedback on the 1-10 kpc scale of galaxy
bulges. Where attempts have been made to characterise the outflows associated
with the high ionisation absorption line systems, the derived masses and mass
outflow rates are often relatively modest (Crenshaw et al. 2003). Fast,
jet-driven outflows of warm, ionised gas have also been detected on a larger
1-100 kpc scale in the extended emission line regions around powerful radio
galaxies (Tadhunter 1991, Clark et al. 1997, Villar Martin et al. 1999) but 
again the masses
involved are relatively small, and their impact on the evolution of the host
galaxy bulges is unclear.

\begin{table*}
\centering
\caption{Parameters of the broad \HI\ absorption}
\begin{tabular}{lrrccrr}
\hline\hline
\strut
Name &  V$_{sys}$ & $\tau$  & N$_{\rm \HI}^a$      & FWZI  & $\dot{M}^b$  & r$_*$  \\
     &   \kms\    &         & 10$^{21}$ cm$^{-2}$   & \kms\ &  M$_\odot$
yr$^{-1}$ & kpc \\
\hline
3C~236 & 30129$^c$     & 0.0033  &   5                   & $-1500$       &  47 & 0.5 \\ 
3C~293 & 13450$^d$     & 0.0038  &   6                   & $-1000(+400)$ &  56 & 1 \\ 
3C~305 & 12550$^e$     & 0.0023  &   2                   & $-500(+150)$  &  12 & 1 \\ 
3C~459 & 66036$^c$     & 0.0005  &   0.75                & $-600(+140)$  &  5.5 & 1 \\ 
4C~12.503 & 36522$^f$  & 0.0017  &   2.6                 & $-1200(+400)$ &  8-21 & 0.02-0.2 \\ 
OQ208  & 22985$^g$     & 0.0057  &   8.3                 & $-1200(+600)$ &  1.2 & 0.01 \\ 
IC~5063 & 3400$^h$     & 0.0120  &   10                  & $-700(+100)$  &  35 & 0.4 \\ 
\hline\hline
\end{tabular}

$a)$ T$_{spin} = 1000$ K;
$b)$ to estimate the mass outflow rate we used the FWZI/2 of the
blueshifted component;
$c)$ Holt (2005),$d)$ Emonts et al. (2005), $e)$ Morganti et al. (2005), $f)$
Holt et al. (2003), $g)$ Marziani et al. (1993), $h)$ Morganti et al. (1998)

\end{table*}

Here we report the detection of several fast, massive outflows of atomic
neutral hydrogen in radio-loud AGN. The detection of such outflows has been
obtained using the unique combination of high sensitivity at 21-cm radio
wavelengths and relatively broad bandwidth now available with the upgraded
Westerbork Synthesis Radio Telescope (WSRT). We further discuss the impact
that they have on the evolution of the host galaxies.

\section{HI outflows: Observations and Results}

We have used the WSRT to observe a small sample of 11 objects characterised by
the presence of a rich ISM surrounding the AGN, e.g. strong CO or far-IR
emission, and/or  known to have undergone a major episode of star formation
in the recent past (Tadhunter et al. 2005) or, as in the case of OQ208, where a
Compton-thick medium is pervading the nuclear environment (Guainazzi et
al. 2004).  Some of them have strong, steep-spectrum core emission (on a scale
$<$10 kpc, unresolved at the resolution of the WSRT 21-cm observations). These
objects are considered to be young or recently restarted radio sources.

Unlike previous \HI\ observations of these galaxies, a wider observing band
(20 MHz and 1024 channels) was used.  The broad instantaneous band covers $\pm
2000$ \kms\ around the central velocity corresponding to the frequency of the
redshifted \HI. This provides a large velocity coverage with sufficient
spectral resolution at both sides of the absorptions, important in order to
obtain a good subtraction of the continuum.  Particular attention has been
also paid to the stability of the instrumental spectral response during the
observations (typically 12 h long). The data were calibrated and reduced using
the MIRIAD package. The final rms noise is between 0.4 and 0.8 \mJybeam\ and
the velocity resolution is about 10\kms.  In each case, the systemic velocity
was derived from recent optical spectroscopy observations (see Table~1) 
and it is indicated in Fig.~1
by a vertical line. The presence of relatively narrow (FWHM $< 200$ \kms) and
deep \HI\ absorption was already known for most of the sample objects, but the
use of the broad-band has allowed the discovery of the much broader and
shallower component of \HI\ absorption.

In all but one (3C~213.1) of the sources in which an optical depth of a few
times 10$^{-4}$ could be reached (i.e.  the sources with radio core flux $\sim
1$ Jy or larger), broad blueshifted \HI\ absorption has been detected.  The
\HI\ absorption profiles of the six detected galaxies are shown in Fig.  1. 
The figure also includes, for completeness, the lower luminosity radio-loud
AGN IC~5063 (PKS~2048-57), the first example in which such broad \HI\ was
detected (Morganti et al.  1997, Oosterloo et al.  2000) using previous
observations with the Australia Telescope Compact Array and LBA.  In the
case of 3C~213.1, we derive a 3-$\sigma$ upper limit for the optical depth of
$\tau = 0.0012$, lower than the typical optical depth of the broad absorption
listed in Table~1.  The remaining three sources (3C~285, 3C~321, 3C~405) of
our initial sample have much weaker radio cores (i.e.  an optical depth of a
few times 10$^{-4}$ could not be reached), therefore it is not too surprising
that no broad \HI\ absorption was detected in these objects.  A description of
the observation and results for the entire sample will be presented in a
forthcoming paper.

\begin{figure*}

\centerline{\psfig{figure=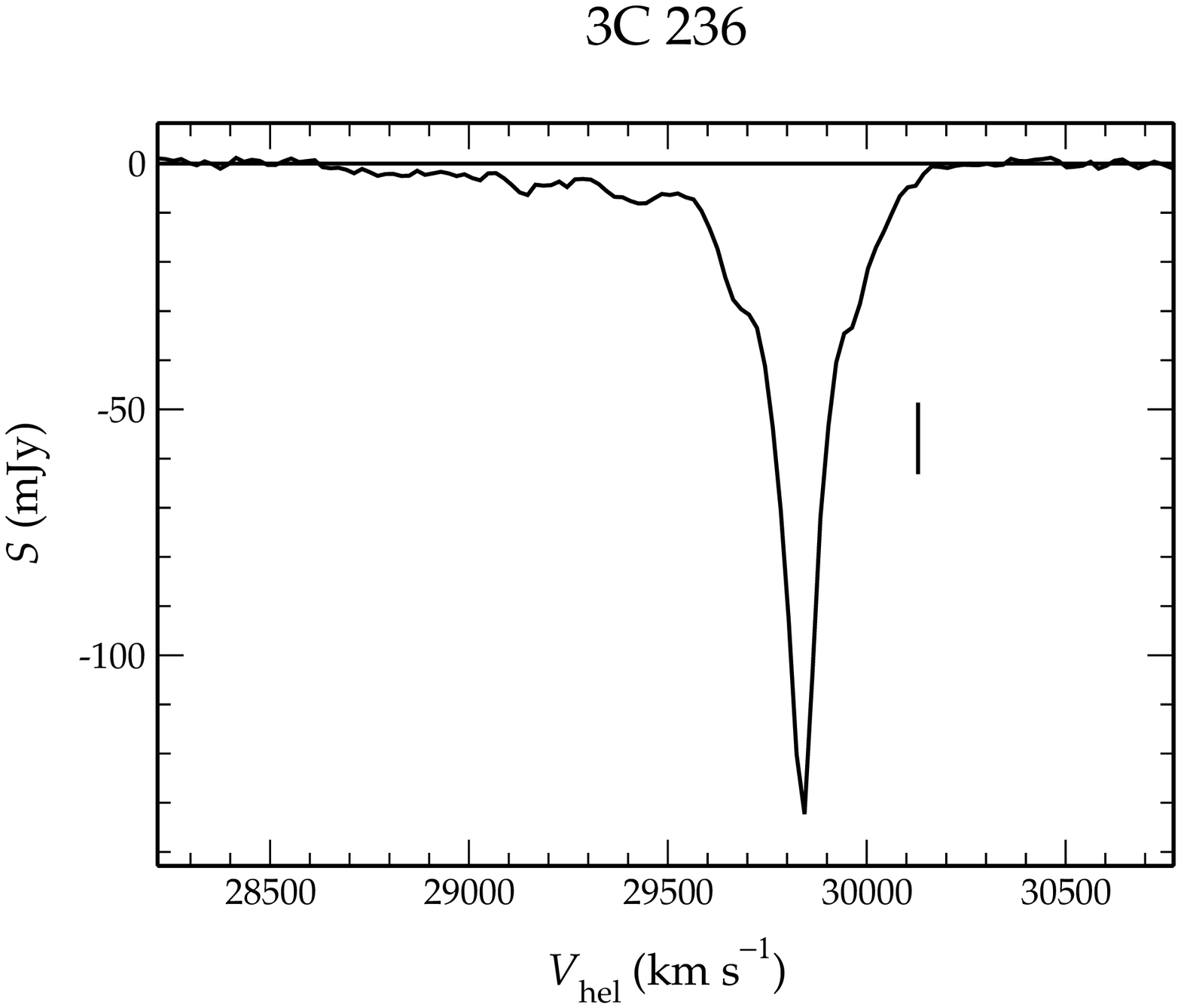,angle=0,height=4.5cm}
\psfig{figure=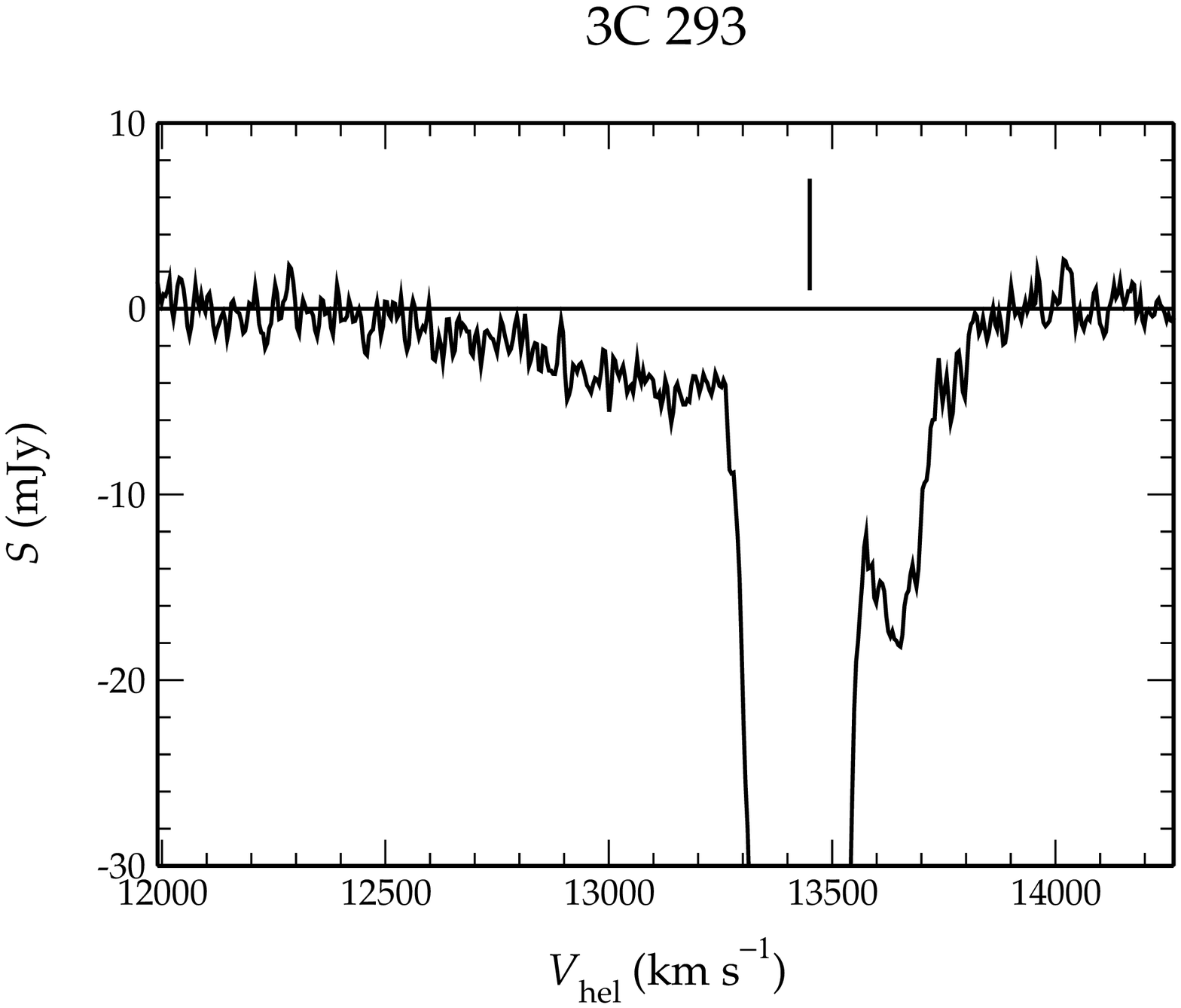,angle=0,height=4.5cm}
\psfig{figure=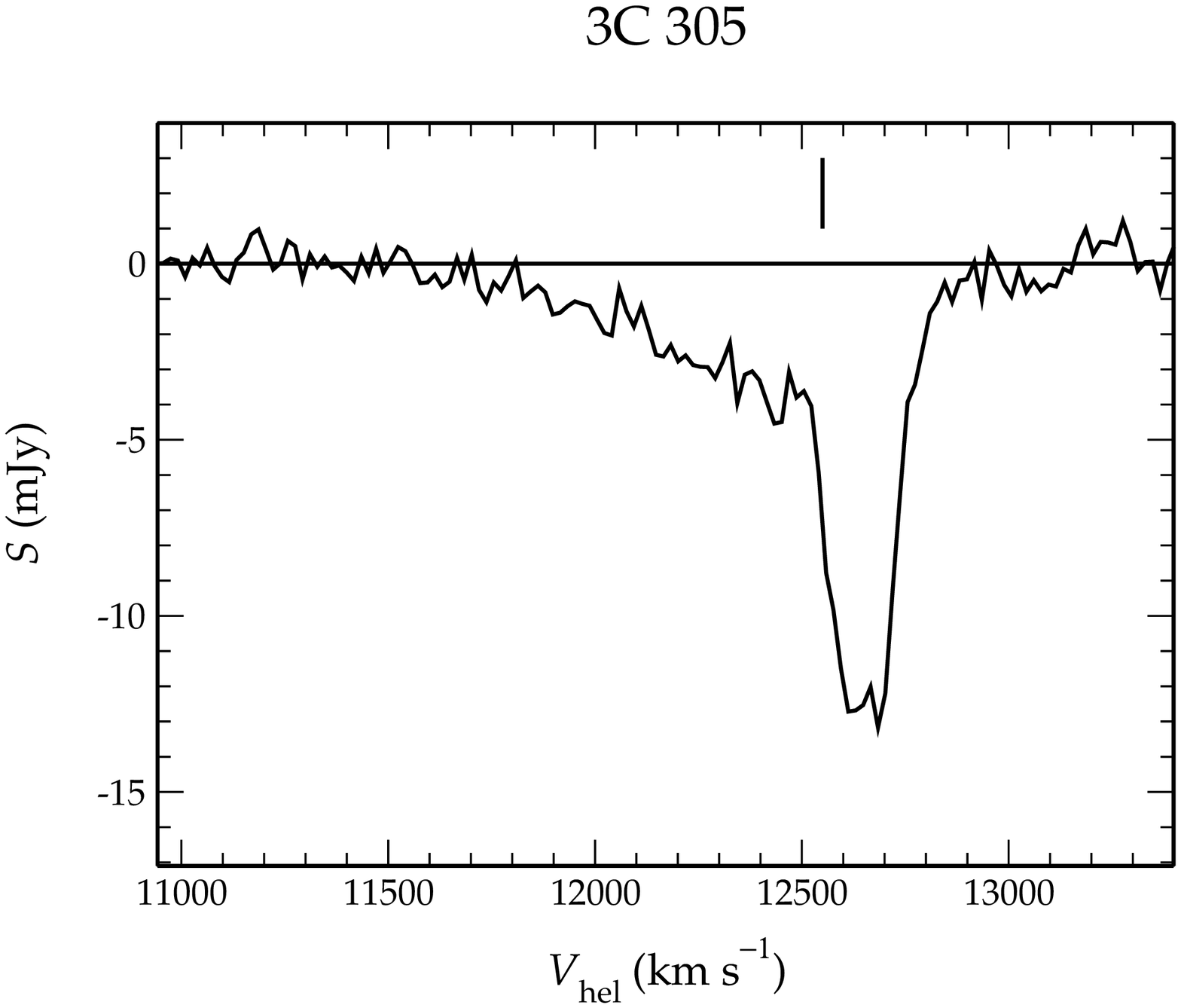,angle=0,height=4.5cm}}
\medskip
\centerline{\psfig{figure=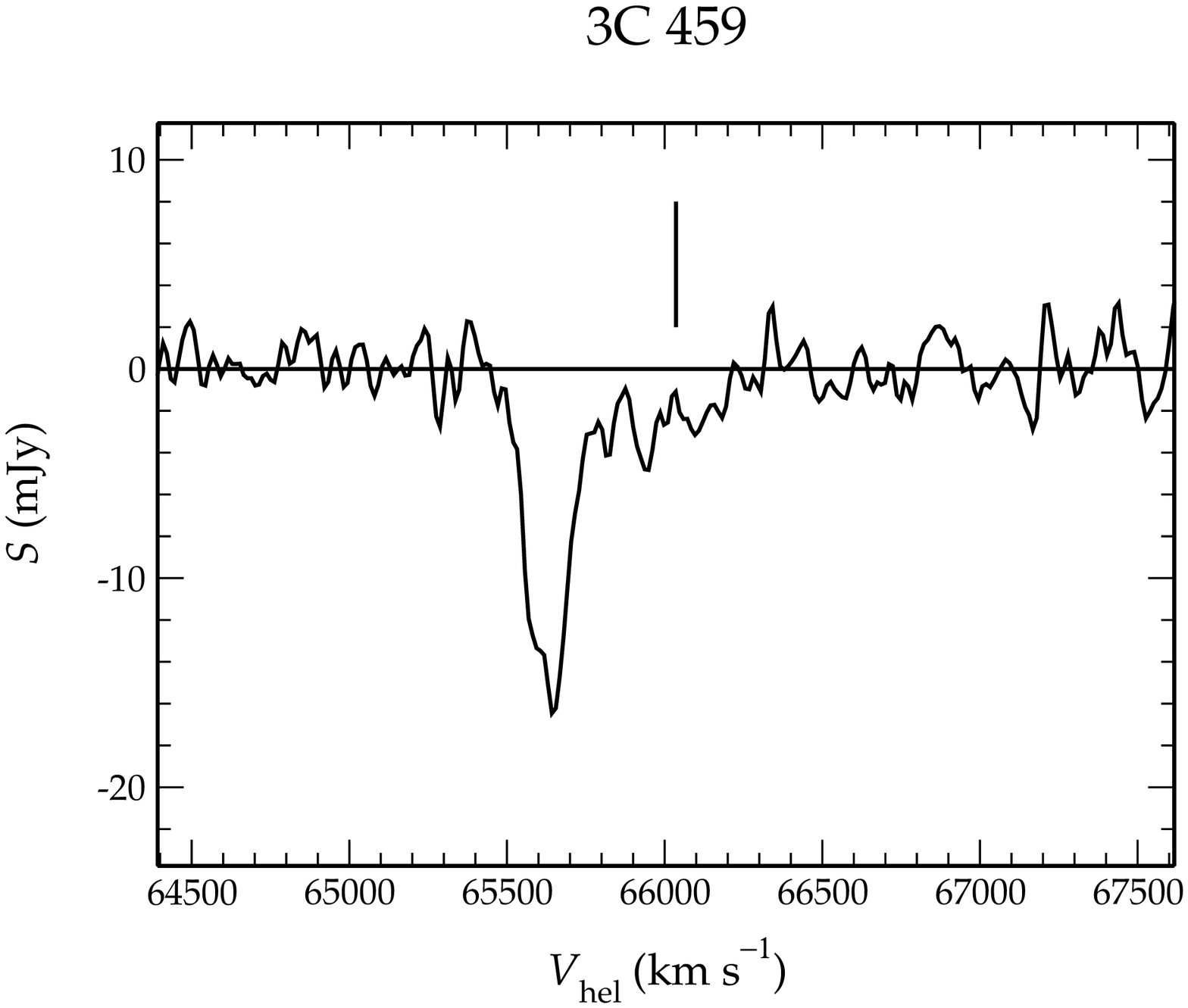,angle=0,height=4.5cm}
\psfig{figure=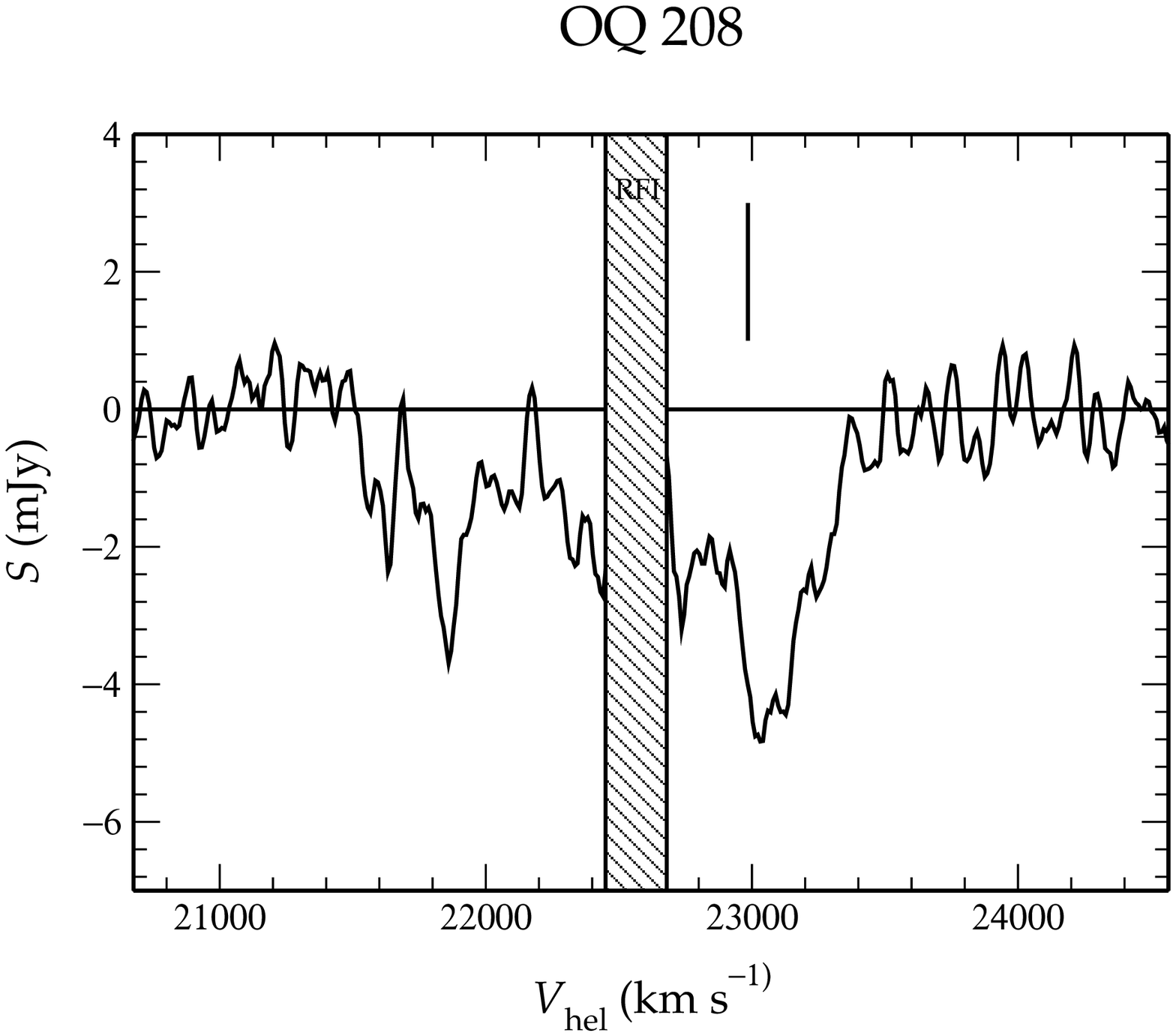,angle=0,height=4.5cm}
\psfig{figure=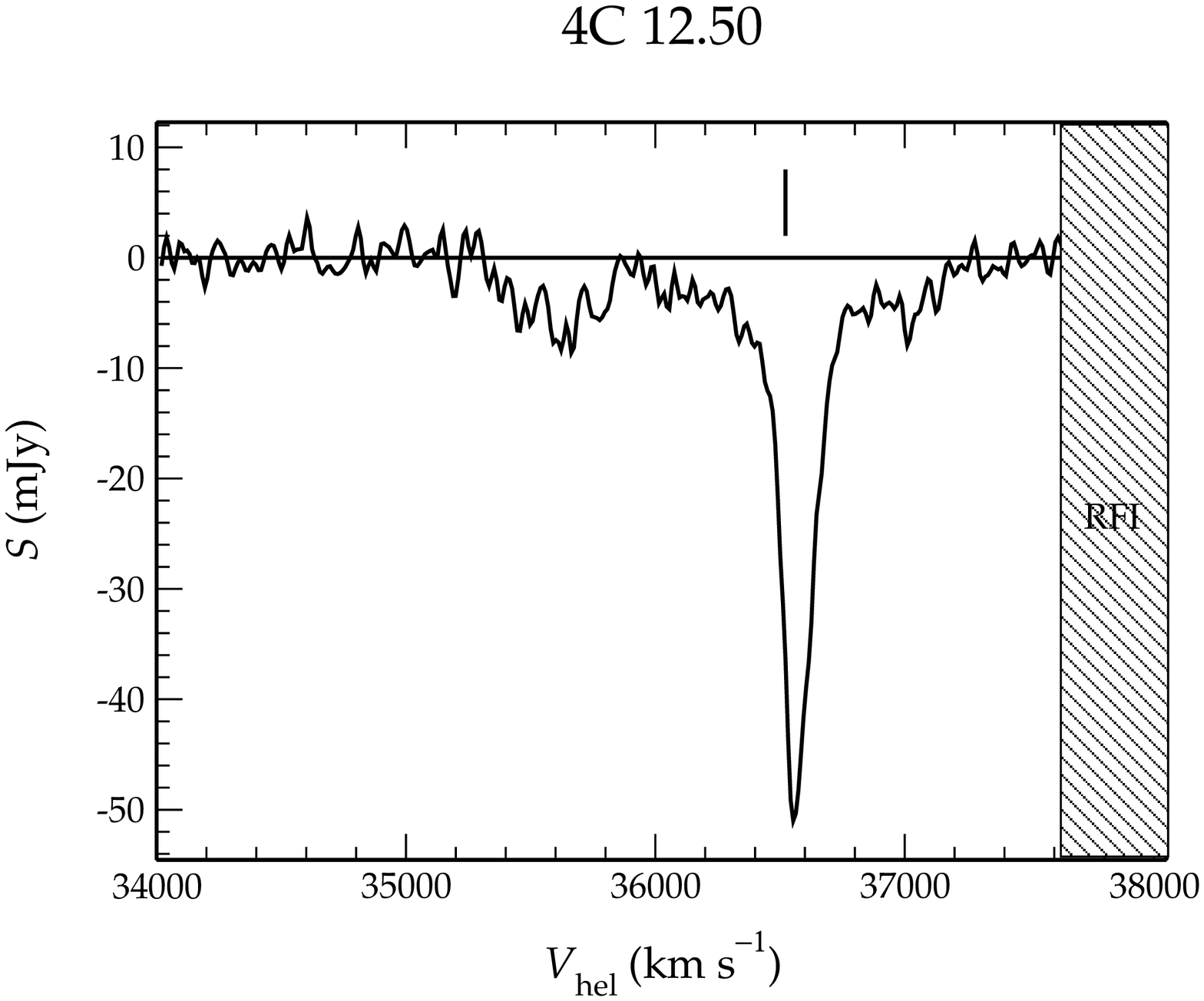,angle=0,height=4.5cm}}
\medskip
\centerline{\psfig{figure=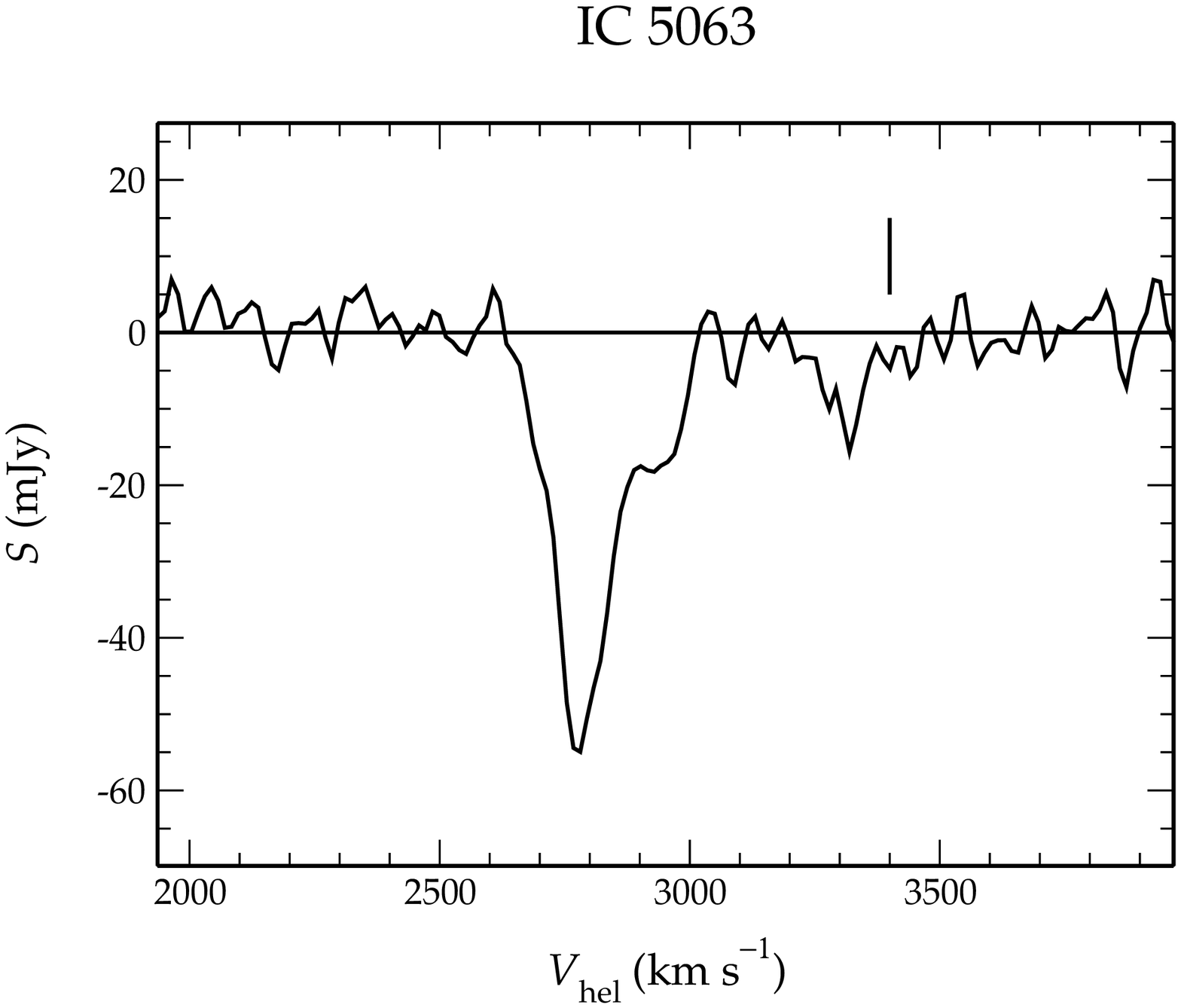,angle=0,height=4.5cm}}
\caption{ 21 cm - \HI\ absorption profiles detected against seven radio-loud
AGN. The observations of the radio galaxies were done using the upgraded
WSRT. The broad \HI\ absorption in the lower luminosity radio-loud AGN IC 5063 
(PKS~2048-57) (detected using the LBA, Oosterloo et al. 2000) is shown for
completeness. The vertical line indicates the systemic velocity.  }
\end{figure*}

The characteristics of the broad \HI\ absorption detected are summarized in Table 1.
The broad \HI\ component is typically shallow (peak optical depth as low as
$\tau \sim 0.0005$), with the full widths at zero intensity (FWZI) of the
absorption ranging between 600 \kms\ and almost 2000 \kms\ --- perhaps the
broadest found so far in \HI\ 21-cm absorption in any class of astronomical
objects. A large fraction of the \HI\ absorption is blueshifted relative to
the systemic velocity of the galaxy. Because the gas producing the absorption
must be in front of the radio source, this provides unambiguous evidence that
the bulk of gas is outflowing.
The typical column densities found for the broad absorption are $1-10 \times
10^{21}$ cm$^{-2}$ (see Table~1). These values for the column density have
been estimated using T$_{\rm spin} =1000$ K. This value is likely to be this
high (or even higher,  Tadhunter et al. in prep.) because of the extreme conditions of the gas, the
proximity to the active nucleus (Bahcall \& Ekers 1969) and the presence of
shocks. The FWZI are given with respect to the systemic velocity. It is clear
that most of the \HI\ absorption is blueshifted, although a redshifted component
(given in parenthesis) is also observed in almost all the objects. In some
cases we know that this component is coming from regularly rotating gas in the
host galaxies (Morganti et al. 2005).

The low optical depth of the broad \HI\ absorption features makes them difficult
to detect using present day radio telescopes, even if they have the
sensitivity and bandwidth of the WSRT. However, the high detection rate in the
selected objects suggests that the broad \HI\ phenomenon is not exceptional,
although it might be connected to a particular stage in the evolution of the
galaxy, for example, the early phases of the expansion of the radio jets
through the bulges of the host galaxies. Indeed, it is intriguing that all the
fast \HI\ outflows have been detected so far are in radio galaxies that are in
the initial (or restarted) phase of their radio activity.

\section{Origin and mass outflow rates}

Different mechanisms have been proposed to explain AGN-driven outflows
(Tadhunter et al. 2001, Elvis et al. 2002, Dopita et al. 2002). In the case of
the fast \HI\ outflows presented here, there is now strong evidence that the
most likely mechanism is the interaction between radio jets and the
surrounding inter stellar medium.  This evidence comprises follow-up
observations at higher resolution of two of the objects (IC~5063 and 3C~305)
in our sample that demonstrate that the outflow regions are resolved on a
scale of 200~pc -- 1.6~kpc, and spatially associated with both the extended
radio knots along the jets and the warm gas outflows detected at the same
locations in optical observations (Oosterloo et al. 2000, Morganti et
al. 2005).  In a third case (3C~293), the evidence that the outflow is located at $\sim
1$ kpc is more indirect (Morganti et al. 2003, Emonts et al. 2005). 
The presence of neutral atomic gas accelerated to such high velocities is in
itself intriguing. It indicates that after a strong jet-cloud interaction the
gas can cool very efficiently. Such rapid cooling is indeed predicted by
recent numerical simulations of jets impacting on gas clouds (Mellema et
al. 2002, Fragile et al. 2004). 
Fast outflows of ionised gas are also detected using optical emission lines in
all the galaxies in Table 1 for which good optical spectra are available (Holt
et al. 2003, 2005; Morganti et al. 2005; Emonts et al. 2005). The presence of
ionised gas can be explained quite naturally as result of the shock produced
by the radio plasma interacting with the ISM.
The similarities found between the kinematics of the neutral and ionised gas
indicate that the two phases of the gas are part of the same outflow.
However, the outflows of ionised gas are typically much less massive (Morganti
et al. 2005, Emonts et al. 2005) than those of the neutral hydrogen.
The mass outflow rate has been
calculated following Heckman (2002) and Rupke et al. (2002):
\begin{equation}
\dot{M} = 30\cdot  {{\Omega}\over{4\pi}}\cdot {{r_*}\over{\rm 1\, kpc}}\cdot
{N_{\rm H}\over{10^{21}\, {\rm cm}^{-2}}}\cdot { v \over 300\, {\rm km
s}^{-1}} \ M_\odot\, {\rm yr}^{-1}
\end{equation}

where the mass is flowing at a velocity $v$ (for which we used the FWZI/2 of
the blueshifted component) into a solid angle $\Omega$ assumed to be $\pi$
steradians from a radius r$_*$, indicated in the last column.  These values
have been obtained for three galaxies from high resolution follow-up
observations (Oosterloo et al.  2000, Morganti et al.  2005, Emonts et al. 
2005).  For the others sources we assumed that the broad absorption comes from
the region with the stronger radio continuum that, in these objects, is
usually not coincident with the radio core, but with blobs along the radio jet
or lobe-like structures.  One should also take into account that the \HI\
absorption can be traced only where the background continuum is observed,
therefore the region of the outflow may be much larger. 

The derived values, ranging from $\sim 1$ to over 50 M$_\odot$ yr$^{-1}$ (see
Table 1), are remarkable.  They are higher than most of the mass outflow
rates deduced for ionized gas using X-ray and UV absorption line measurements
of nearby AGN (Crewshaw et al.  2003), although in some nearby Seyfert
galaxies value up to 10 M$_\odot$ yr$^{-1}$ have been also derived (Veilleux
et al.  2005).  The values we derive for the neutral hydrogen outflows,
correspond to the lower end of the distribution for starburst super-winds in
Ultra-Luminous IR Galaxies (ULIG) that show outflows rates (Heckman 2002,
Rupke et al.  2002, 2005a,b) ranging from 10 to 1000 M$_\odot$
yr$^{-1}$. 

\section{Implications for the evolution of the host galaxies}

Starburst super-winds are recognized to be responsible for inhibiting early
star formation, enriching the ICM with metals and heating the ISM/IGM medium. 
Given the mass outflow rates obtained above, the results presented here show
that similar effects could be produced by the jet during the radio-loud phase
of AGN activity.  Indeed, the jet in powerful radio galaxies (as those
considered in this work) can be extremely energetic (\gtsima 10$^{58}$ erg,
when integrated over the lifetime of the radio source), often exceeding the
gravitational binding energy of the gas in a typical galaxy, and only a
fraction of the power of the jets is needed to produce the \HI\ outflows. 

More in general, to investigate the importance of the \HI\ outflows, we can
compare the the bulk energy of the outflow with the typical binding energy of
the gas in the galaxy.  Assuming a typical galaxy with a gas mass of M$_{\rm
gas}\sim 1 \times 10^{10}$ M$_\odot$, at a typical radius of $r_g \sim 5$ kpc,
and a total mass within 5 kpc of M$_{\rm gal} \sim 10^{11}$ M$_\odot$, the
binding energy is $\sim 2 \times 10^{58}$ erg.  For comparison, a 10 M$_\odot$
yr$^{-1}$ outflow moving outwards at 500 \kms\ has a bulk kinetic energy of
$\sim 2.5 \times 10^{57}$ erg (assuming that the outflow is integrated over
$10^8$ yr lifetime of the radio source).  This value could easily be increased
by a factor of two if we take into account the turbulent motion (indeed the
\HI\ absorption lines are as broad as they are shifted).  Thus, the kinetic
power in the \HI\ outflows can represent a substantial fraction of the energy
required to remove the gas from the bulges of the galaxies, and, therefore,
can provide an important source of feedback. 

The massive \HI\ outflows show, more clearly than optical emission line studies
(Tadhunter et al. 2001), that AGN have a major effect on the circumnuclear
gas; the central kiloparsec region of AGN host galaxies are clearly dynamic
places with the distribution of ISM evolving strongly as a function of time.
Our results also have implications for the evolution of the radio
sources. Although, as mentioned above, the radio jets in these galaxies are
unlikely to be completely confined by the surrounding dense interstellar
medium, the strong interaction that we detect through the kinematic signature
of the broad \HI\ can disrupt the paths of the jets and slow down their overall
expansion, as predicted in numerical simulations (Morganti et al. 2004, Saxon
et al. 2005). This effect is likely to be even more relevant in higher-density
conditions prevailing in the early universe.

To date, the majority of studies
of atomic neutral hydrogen in radio galaxies have been made using a relatively
narrow bandwidth and low sensitivity. This automatically limits the
possibility of detecting \HI\ with the extreme kinematics described here.  
Although the detection of this component is complicated by the low optical
depth, future observations extended to larger samples of both compact and more
extended radio galaxies will reveal whether this phenomenon is significant at
all stages of evolution of powerful, radio-loud AGN.

\begin{acknowledgements}

The Westerbork Synthesis Radio telescope is operated by the ASTRON (Netherlands
Foundation for Research in Astronomy) with support of the Netherlands
Foundation for Scientific Research (NWO).

\end{acknowledgements}

\end{document}